\documentclass[intlimits,twoside,a4paper]{article}

\usepackage[cp1251]{inputenc}
\usepackage[eqsecnum]{cmpj3}
\usepackage{graphicx}

\usepackage{bm}
\usepackage{multirow}

\issue{2022}{25}{2}{23301}
\doinumber{10.5488/CMP.25.23301}

\title[Dynamics of Bose--Einstein condensates under anharmonic trap]%
{Dynamics of Bose--Einstein condensates under anharmonic trap}%

\author[H. Al-Jibbouri]{H. Al-Jibbouri\orcid{0000-0001-8429-9158}}
\address{College of Science, University of Al-Qadisiyah, Al-Diwaniya 58002, Al-Qadisiyah, Iraq}

\date{Received February 11, 2022, in final form April 21, 2022}

\begin{document}

\maketitle

\begin{abstract}
The dynamics of weakly interacting three-dimensional Bose--Einstein condensates (BECs), trapped in external axially symmetric plus anharmonic distortion potential are studied. Within a variational approach and time-dependent Gross--Pitaevskii equation, the coupled condensate width equations are derived. By modulating anharmonic distortion of the trapping potential, nonlinear features are studied numerically and illustrated analytically, such as mode coupling of oscillation modes, and resonances. Furthermore, the stability of attractive interaction BEC in both repulsive and attractive anharmonic distortion is examined. We demonstrate that a small repulsive and attractive anharmonic distortion is effective in reducing (extending) the condensate stability region since it decreases (increases) the critical number of atoms in the trapping potential.
%
\end{abstract}

\section{Introduction}

The successful detection~\cite{BECs1,BECs2,BECs3,BECs4,BECs5,BECs6,BECs7} of Bose--Einstein condensates (BECs) at ultra-low temperatures plays a major role in quantum many-body physics, and introduces several theoretical researches on different features of the condensate~\cite{theo1,theo2,theo3,theo4,theo5,theo6,theo7,theo8,theo9}. The features of a hypothetical condensate at low temperature and the neglection of all correlations between the atoms are basically defined by a underlying time-dependent Gross--Pitaevskii (TDGP) equation~\cite{GP1,GP2} that sufficiently combines the trapping potential and the interaction between the particles that produce the condensate. The impact is intrinsically quantum mechanical if we imagine a set of particles in equilibrium state at a certain temperature $T$. In the quantum world, any particle is performed by a focused wavepacket, with its spatial range existence described by the de Broglie wavelength. This wavelength might become equivalent to the average distance between the particles at low temperatures and/or high particle densities.

Indeed, the majority of BEC research has focused on a condensation trapped in a harmonic (parabolic) trap potential due to the large size of boosting magnetic field elements, for instance,~\cite{5,55}. One of the most exciting features is the condensate's collective excitation produced by a dramatic change in the contact interaction. The dynamics behavior of the condensate under a rapid or continuous modification of the trapping frequency was studied experimentally~\cite{5,55,555} and theoretically~\cite{6,66}. Similar studies were performed by modifying the anisotropy of the confining harmonic potential~\cite{cvc,Hamid1,Hamid2} or by adjusting the atomic scattering length~\cite{7,IVidanovic}. Within the precise underlying Gross--Pitaevskii (GP) framework, the dynamics of a BEC in an anharmonic potential was studied explicitly in~\cite{ohtt,an3}. Indeed, the oscillation modes of a 1D BEC trapped in anharmonic potential were discussed with a system of two-body interaction ~\cite{10l,11b}, two-body and three-body interaction at a low temperature~\cite{11c,10e}, as well as for finite temperature~\cite{11}, and dipolar Bose gas in~\cite{an6,2017,2019,2020}. The oscillation modes of a 3D BEC in an anharmonic trap were studied in~\cite{aljibbouri+an}. In~\cite{fermi}, the eigenmode frequencies of trapped anharmonically superfluid Fermi gas were investigated in the transition from a Bardeen--Cooper--Schrieffer to a Bose--Einstein condensate (BCS-BEC).
BECs in systems with attractive interactions were understood and observed experimentally with ${}^7$Li in~\cite{BECs5,BECs6}. The sign and magnitude of the interaction between atoms can be set to any value: large
or small, repulsive or attractive. The existence of an attractive
interactions between the atoms influences
the stability region of a BEC, where the condensate becomes unstable and collapses for large enough attractive interaction strength~\cite{7,4,theo8,60}.
The critical number of particles was obtained numerically~\cite{5,6,6aa,6a} by using Gross--Pitaevskii formalism in cylindrical traps. References~\cite{11c,10} studied how much the small anharmonic distortion added to the confining potential affected on the critical number of particles with negative interatomic interaction strength or two- and three-interaction strengths, respectively.
The stability of attractive BEC was studied in the presence of a repulsive three-body interaction ~\cite{Hamid2}.

Motivated by this, the dynamics and low-energy oscillation modes BECs of elongated anharmonic potential is discussed at ultra-low temperature. Within variational analysis of GP equation, coupled condensate width equations were obtained and derived. By using linear response theory, the low-energy oscillation modes are studied, where the nonlinearity between modes is mentioned. The stability of attractive interaction BEC in both repulsive (attractive) anharmonic distortion gives rise to a destabilisation (stabilization) of the condensate.

The paper is prepared as follows. The analysis of obtaining the equation of motion of the condensate widths in axially symmetric plus anharmonic distortion potential as well as the eigenmode frequencies is presented in section~\ref{sec:method},. In section~\ref{sec:RS}, the periodic oscillation due to the anharmonic elongated trap potential is discussed. In section~\ref{sec:stability}, the stability diagram of attractive interaction strength with repulsive (attractive) anharmonic distortion is discussed. Finally, section~\ref{sec:con} presents our conclusions.

\section{Method}
\label{sec:method}

At zero temperature, the dynamics of atomic BEC were mentioned by $3$-dimensional TDGP equation~\cite{GP1,GP2}
\begin{equation}
\ri\hbar\frac{\partial}{\partial t}\Psi({\bf r},t)=\left[-\frac{\hbar^{2}}{2
M}\nabla^2+V_{\rm trap}({\bf r}) + g \left|\Psi({\bf r},t)\right|^{2}
\right]\Psi({\bf r},t)\,.
\label{eq:GP}
\end{equation}
Here, $\Psi({\bf r},t)$ refers to the wave function. The first term in GP equation is kinetic energy, while other terms are an external axially symmetric plus anharmonic potential $V_{\rm trap}({\bf r})=\frac{1}{2}M \omega_{\rho}^2 \left(\rho^2+ \lambda^2 z^2\right)+ \delta \rho^4$
with anisotropy $\lambda$, anharmonic parameter $\delta$, oscillator frequency $\omega_{\rho}=\omega_x=\omega_y$, $\rho^2=x^2+y^2$, and nonlinear term arising due to the effective
contact interatomic interaction $g=4 \piup N a/M$ that is coupled with the scattering-length as well as the total amount particles in the condensate $N$. Either GP equation is solved numerically~\cite{12} or it is solved by
considering a variational approach of the Gaussian form~\cite{Perez1,Perez2}. In the latter case, the variational problem represents the extremization of the action defined by the Lagrangian function of equation~(\ref{eq:GP})
\begin{eqnarray}
\hspace{-1.5cm} L(t)&=&\int \, \rd {\bf r} \bigg[- \frac{\hbar^2}{2m}|\nabla
\Psi|^2-V_{\rm trap}({\bf r}) |\Psi|^2  -
\frac{g}{2} |\Psi|^4  +\,\frac{\ri \hbar}{2}\left(\Psi^* \frac{\partial
\Psi}{\partial
t}- \Psi\frac{\partial \Psi^*}{\partial t} \right)\bigg]  \, .
\label{eq:lagrange}
\end{eqnarray}
The dynamics of BEC system can be studied analytically by using the Gaussian variational ansatz~\cite{Perez1,Perez2}.
\begin{equation}
\hspace{-0.15cm} \Psi(\rho,z,t)=N(t) \exp \left\{\sum_{r=\rho,z} \left[\left(-\frac{1}{2\eta_{r}^2}-\ri
 \alpha_{r}\right)r^2\right]\right\},
\label{eq:gaussainansatz}
\end{equation}
with the normalization wave function $N(t)=1/\sqrt{\piup^{{3}/{2}}
\eta_{\rho}^2 \eta_z}$. $\eta_{r}$  and $\alpha_{r}$ are variational parameters, clarifying
the condensate widths and phases, respectively. Substituting equation~(\ref{eq:gaussainansatz}) into equation~(\ref{eq:lagrange}), we obtain
\begin{eqnarray}
\hspace{-1.5cm} L(t)&=&\, -\frac{\hbar^2}{2
M}\left(\frac{1}{\eta_{\rho}^2}+4\alpha_{\rho}^2\eta_{\rho}^2+\frac{1
}{2\eta_z^2}+2\alpha_z^2\eta_z^2\right)  -\frac{\hbar}{2}\left(2\dot \beta_{\rho} \eta_{\rho}^2 +\dot \alpha_z
\eta_z^2\right)  \nonumber\\ \hspace{-1.5cm}&-&
\frac{g}{2(2 \piup)^{3/2} \eta_{\rho}^2 \eta_z}  -\frac{M \omega_{\rho}^2}{2}\left(\eta_{\rho}^2+\gamma^2
\frac{\eta_z^2}{2}\right) +2\delta M \omega_{\rho}^2 \eta_{\rho}^{4}\, .
\label{eq:lagf}
\end{eqnarray}
The equations of motion for all variational parameters are derived from the relevant Euler--Lagrange equations. The phases $\alpha_{\rho}$ and $\alpha_z$ may be represented directly in terms of first derivatives of the condensate widths $\eta_{\rho}$  and $\eta_z$, yielding,
\begin{align}
\eta_{\rho,z}& = \frac{M \dot \eta_{\rho,z}}{2 \hbar \eta_{\rho,z}}\, .
\label{eq:phizu}
\end{align}
After providing dimensionless parameters, $\eta_{\rho,z} \rightarrow l (\eta_{\rho,z}),\  t \rightarrow t \omega_{\rho}, \ \delta=\delta/(\hbar \omega_{\rho})$ with the oscillating length $l=\sqrt{\hbar/(M \omega_{\rho})}$ as well as $\mathcal{P}=\sqrt{\piup/2} a N/l$ defining the dimensionless interaction strength with the $s$-wave scattering length $a$, particle number $N$, equation~(\ref{eq:phizu}) is inserted into the Euler--Lagrange equations for the width of the condensates $\eta_{\rho,z}$. A set of second order ordinary differential equations for $\eta_{\rho}$ and
$\eta_z$ in the dimensionless form are demonstrated:
%
\begin{eqnarray}
&& \ddot{\eta}_{\rho}(t)+\eta_{\rho}(t)+8\ \delta \ \eta_{\rho}^{3}(t)-\frac{1}{\eta_{\rho}^3(t)}-\frac{\mathcal{P}}{\eta_{\rho}^3(t) \eta_z(t)}=0 \label{eq:Rr}\,, \\
&& \ddot{\eta}_z(t)+ \lambda^2 \eta_z(t) -\frac{1}{\eta_z^3(t)}-\frac{\mathcal{P}}{\eta_{\rho}^2(t) \eta_z^2(t)}=0\label{eq:Rz}.
\end{eqnarray}
%
For ${}^{87}$Rb BEC atoms~\cite{BECs3}, $M = 1.44 \cdot 10^{ -25}$~kg, $\omega_{\rho}=2\piup \cdot 126$~Hz, $\omega_{z}=2\piup \cdot 21$~Hz,
 $N=1 \cdot 10^5$ atoms, $a=100 a_0$, where $a_0$ is Bohr radius. By applying an external magnetic field through a Feshbach resonance, the amount of interaction strength may be adjusted to any value, large or small, negative or positive. A controllable parameter $\delta$ can be assumed so that the anharmonicity is within a certain range of $\delta\ll 1$.

\begin{figure}[t]
	\centering
	\includegraphics[width=6cm]{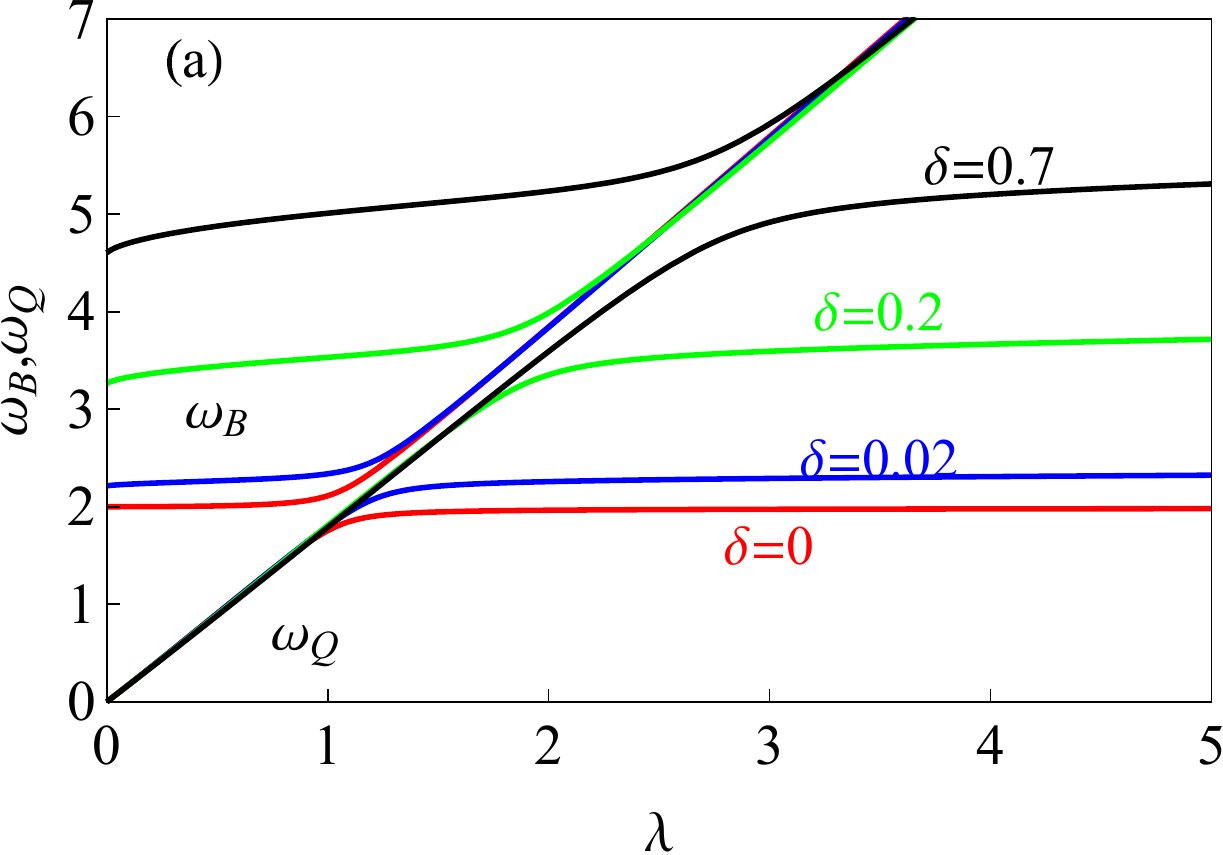}
	\includegraphics[width=6cm]{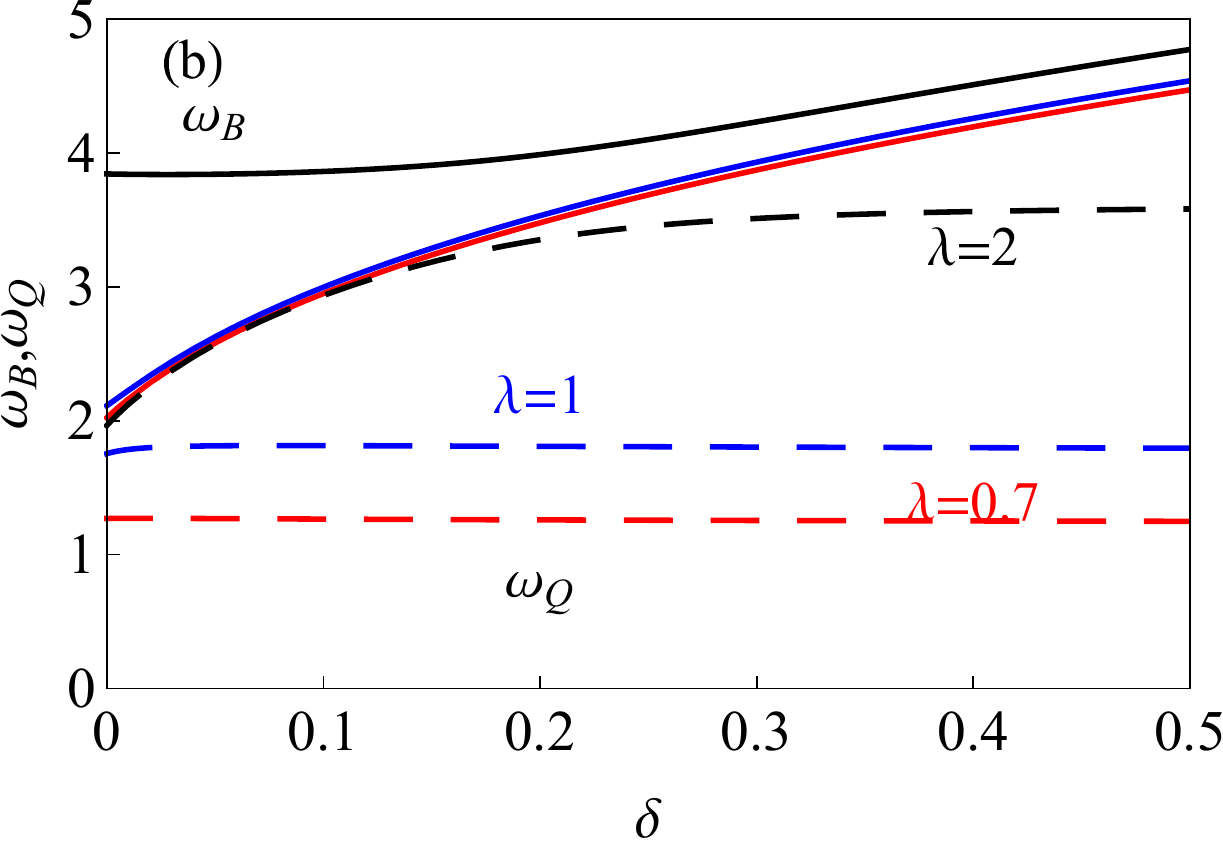}
	\caption{(Colour online) (a) Quadrupole $\omega_{\rm Q}$ and breathing $\omega_{\rm B}$ eigenmode frequencies in the unit of $\omega_{\rho}^{-1}$ versus $\lambda$ for several values of anharmonic distortion $\delta=0,0.02, 0.2, 0.7$ (red, blue, green, black) curves, respectively, and $\mathcal{P}=10$. (b) Quadrupole $\omega_{\rm Q}$ (dotted curves) and breathing $\omega_{\rm B}$ (solid curves) modes in unit of $\omega_{\rho}^{-1}$ versus the anharmonic distortion $\delta$ for different trap anisotropy $\lambda=0.7,1,2$, (red, blue, black) curves, respectively, and $\mathcal{P}=10$.}
	\label{fig:freQB}
\end{figure}

The equilibrium positions $\eta_{\rho_0}$ and $\eta_{z_0}$ are evaluated by putting the acceleration of equations~(\ref{eq:Rr}) and~(\ref{eq:Rz}) to zero [i.e., $\ddot{\eta}_{\rho}(t)=\ddot{\eta}_{z}(t)=0$], we get
\begin{eqnarray}
&& \eta_{\rho_0}+8\ \delta \ \eta_{\rho_0}^{3}-\frac{1}{\eta_{\rho_0}^3}-\frac{\mathcal{P}}{\eta_{\rho_0}^3 \eta_{z_0}}=0 \label{eq:Rr00}\,, \\
&& \lambda^2 \eta_{z_0} -\frac{1}{\eta_{z_0}^3}-\frac{\mathcal{P}}{\eta_{\rho_0}^2 \eta_{z_0}^2}=0\label{eq:Rz00}.
\end{eqnarray}
The analytical result of low-lying collective modes is evaluated formerly using Gaussian approximation~\cite{Perez1,Perez2}. By linearization
equations~(\ref{eq:Rr}) and~(\ref{eq:Rz}) around equations~(\ref{eq:Rr00}) and~(\ref{eq:Rz00}), frequencies of quadrupole $\omega_{\rm Q}$ and breathing $\omega_{\rm B}$ modes were obtained
\begin{eqnarray}
&&\omega_{\rm{B,Q}}=\sqrt{\frac{\alpha_1+\alpha_3 \pm \sqrt{(\alpha_1-\alpha_3)^2+8\alpha_2^2}}{2}}\,,
\label{eq:frequencyBQ}
\end{eqnarray}
where $\alpha_1=1+ 8\ \delta \eta_{\rho_0}^{3}+3/\eta_{\rho_0}^4+3 \mathcal{P}/(\eta_{\rho_0}^4\eta_{z_0}), \ \alpha_2=\mathcal{P}/(\eta_{\rho_0}^3\eta_{z_0}^2)$, and $\alpha_3=\lambda^2+3/\eta_{z_0}^4+2\mathcal{P}/(\eta_{\rho_0}^2 \eta_{z_0}^3)$.
For the repulsive interaction, the breathing and quadrupole modes are characterized in (out)-phase radial and axial oscillations, respectively.
The frequencies of equations~(\ref{eq:frequencyBQ}) substantially rely on the anharmonic distortion $\delta$ and the anisotropy $\lambda$. Figure \ref{fig:freQB}~(a) denotes the quadrupole $\omega_{\rm Q}$ and breathing~$\omega_{\rm B}$ eigenmode frequencies as a function of $\lambda$ with different $\delta=0$, $0.02$, $0.2$, $0.7$ (red, blue, green, black) curves, respectively. It is clear that the eigenmode frequencies change strongly with respect to the anharmonic distortion $\delta$. Figure~\ref{fig:freQB}~(b) refers the quadrupole $\omega_{\rm Q}$ and breathing~$\omega_{\rm B}$ eigenmode frequencies versus the anharmonic distrotion of the trapping potential $\delta$ for $\lambda=0.7$, $1$, $2$ (red, blue, black) curves, respectively. In particular, the frequencies in figure~\ref{fig:freQB}~(b) deviate due to the coherent wave response of the anharmonic distortion potential $\delta$. On the other hand, the condensate widths will broaden due to the atom interaction. However, we read a tiny effect for eigenmode frequencies at small $\lambda$, whereas  the breathing and quadrupole eigenmode frequencies strongly depends on the anharmonic distortion $\delta$ for large $\lambda$. Due to the anharmonicities of the trapping potential,
the eigenmodes deviated from the values estimated via linear stability analysis, equation~(\ref{eq:frequencyBQ}). Furthermore, large oscillations were expected at $\Omega \approx \omega_{\rm{B,Q}}$.

\section{Resonance analysis}
\label{sec:RS}

In the previous works~\cite{IVidanovic}, the nonlinear coupling
emerges from the essential interaction between particles in the
system. Here, we display a nontrivial nonlinear
coupling mechanism by anharmonic corrections to
the trap; for time
dependent $\delta(t)$, yielding
\begin{equation}
\delta(t)=\delta+ q \ \cos(\Omega \ t)\,.
\end{equation}
\label{sec:Resonance}
The dynamics is obtained via the modulation which strongly relies on the driving frequency $\Omega$.

\begin{figure}[htb]
	\centering
	\includegraphics[width=6cm]{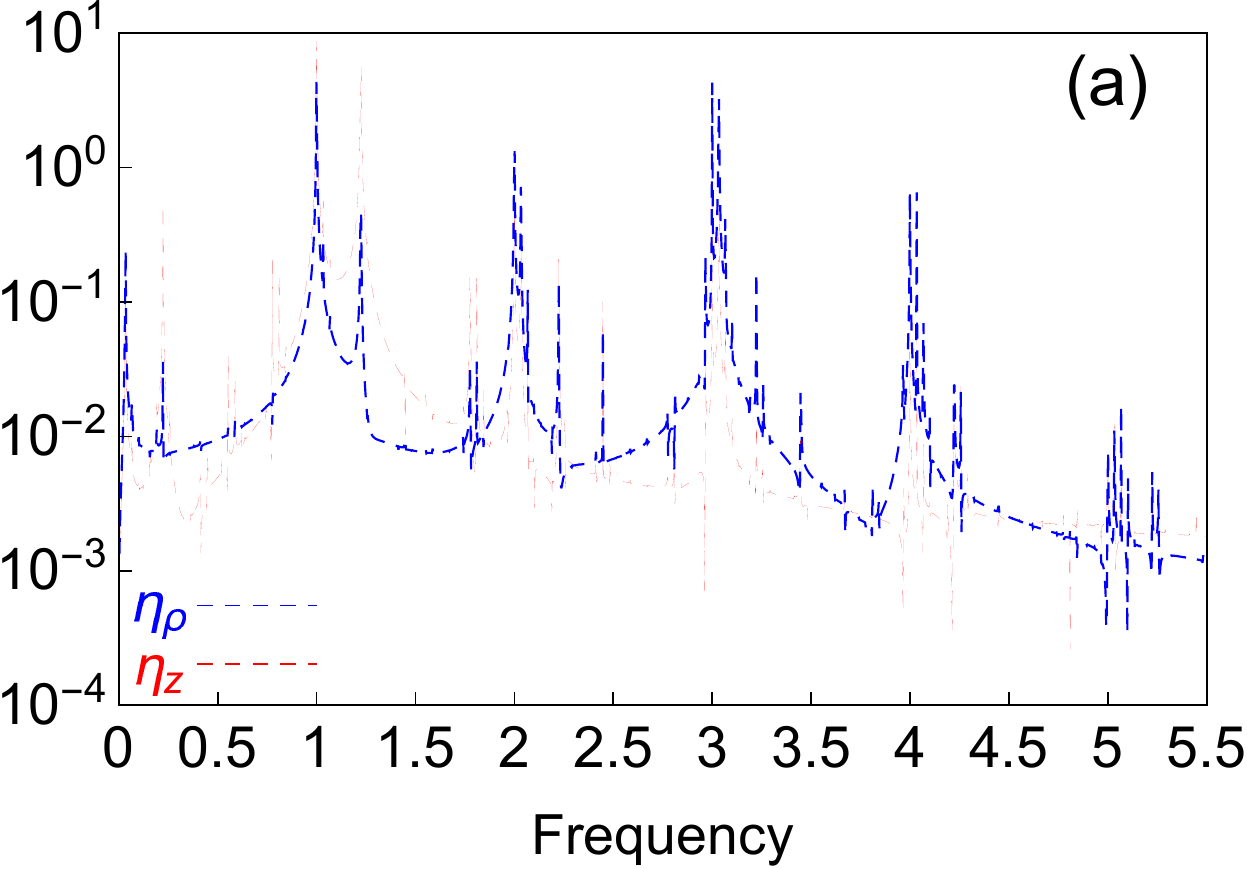}
	\includegraphics[width=6cm]{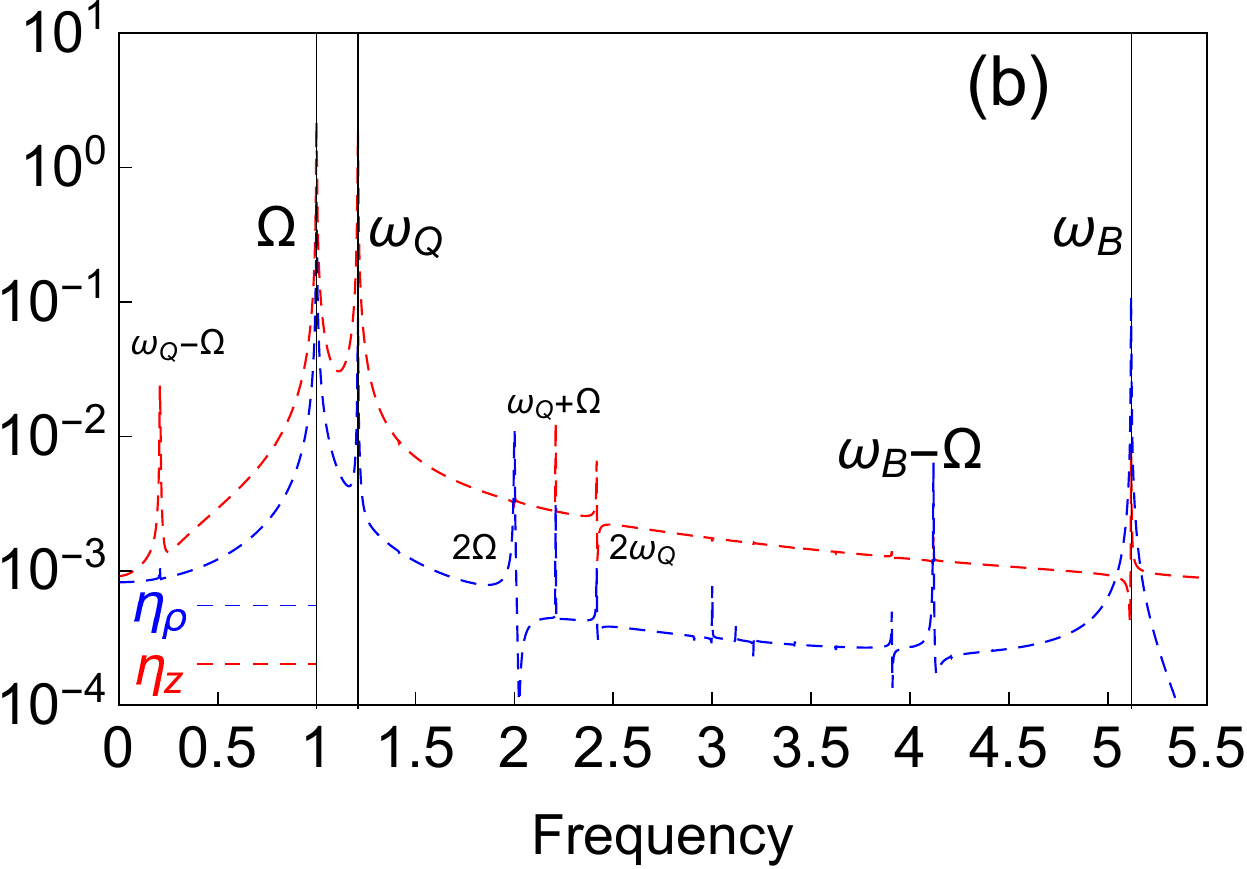}
	\caption{(Colour online) High-resolution Fourier spectrum of the condensate widths $\eta_{\rho}(t)$ and $\eta_{z}(t)$ for $\mathcal{P}=10$ $\lambda=0.7$, $\Omega=1$, $q=0.01$, $\delta=0.1$ (left-hand) and $\delta=0.7$ (right-hand).}
	\label{fig:spectrum}
\end{figure}

Numerical solution for the time interval $(0,T)$ of equations~(\ref{eq:Rr}) and~(\ref{eq:Rz}) is then analyzed using the discrete Fourier transform. The
numerical values were obtained with the time step $\Delta t$ and perform the discrete transformation from time to frequency
representation. The maximal value of $\omega$ accessible this way is given by $\omega_{\rm max}= \piup/\Delta t $. The resolution of the obtained spectrum is determined by the value of $T$ and it reads $\Delta \omega=2\piup/T$. We start by considering high-resolution Fourier spectrum, using $T = 10000$. Furthermore, equations~(\ref{eq:Rr}) and~(\ref{eq:Rz}) solve numerically and obtain the Fourier
transform of there solutions~\cite{Mathematica}. For instance, an excitation spectrum obtains for $\mathcal{P}=10$,
$\lambda=0.7$, $\Omega=2$, as well as $\delta=0.1,0.7$ figure~\ref{fig:spectrum}~(a) and~(b), respectively. On the following graphs, we look at the spectrum part by part and present all illustrious peaks. Using
simple inspection, we find that they correspond to the driving frequency $\Omega$ and to the eigenmode frequencies $\omega_{\rm Q,B}$. Higher-order harmonics  $n\Omega$ and $m\omega_{\rm Q,B}$ are also present in the spectrum, together with the mixed modes $n\Omega+m\omega_{\rm Q,B}$ and the beating frequency $|n\Omega-m\omega_{\rm Q,B}|$ and prominent
modes are more complicated for a small anharmonic distortion figure~\ref{fig:spectrum}~(a).
Resonant effects are expected for $\Omega \approx \omega_{\rm Q,B}$, as numerical results clearly confirm. However, beside the expected basic modes there is a complex peak structure, figure~\ref{fig:spectrum}. Here, we explain its origin. As already stated, our spectrum contains coupled modes of the form $n\Omega+m\omega_{\rm Q,B}$. As a consequence of nonlinearity and anharmonicity, distortion in our basic
equations, more general form $n\Omega+m\omega'_{\rm Q,B}$ is assumed, where $\omega'_{\rm Q,B}$ denotes the eigenmodes of collective oscillations close to
a resonance. Far from the resonance, we have $\omega'_{\rm Q,B} \approx \omega_{\rm Q,B}$, but as we approach the resonance regime, $\omega'_{\rm Q,B}$ shifts from $\omega_{\rm Q,B}$ and
needs to be determined without relying on linear response theory.

\begin{figure}[htb]
	\centering
	\includegraphics[width=6cm]{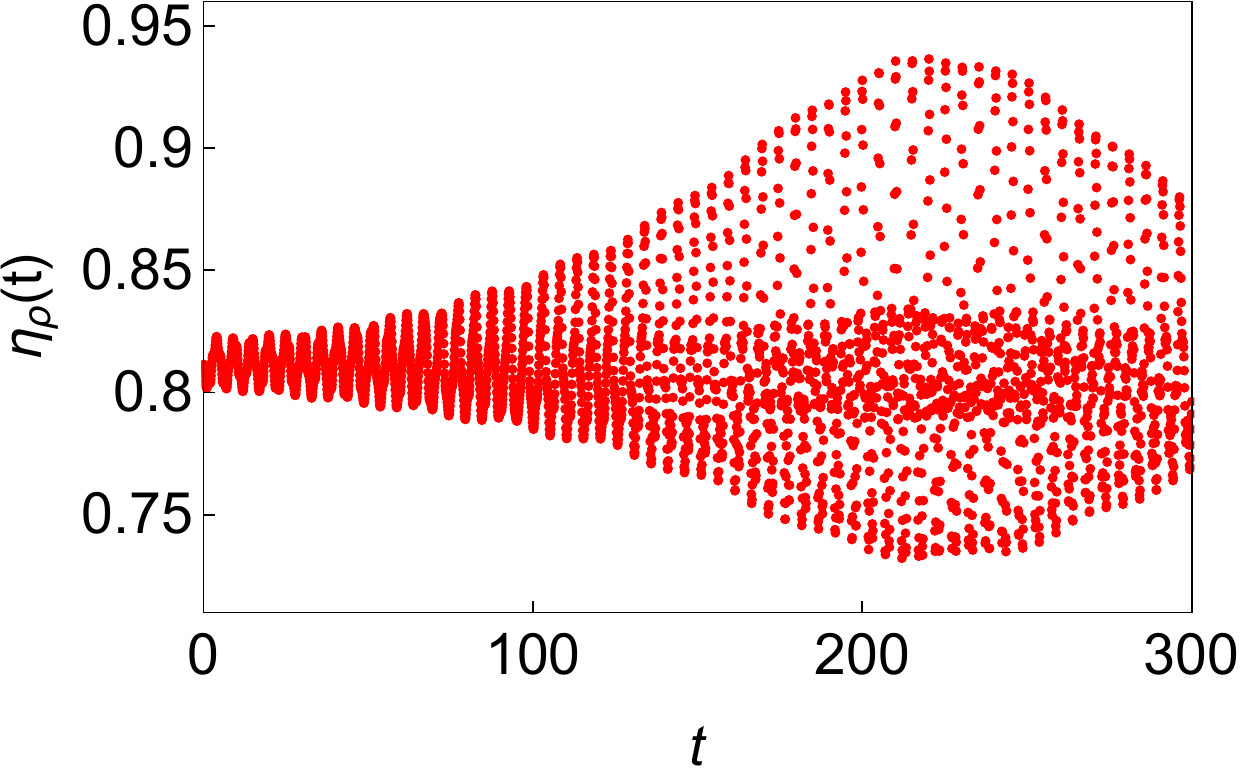}
	\includegraphics[width=6cm]{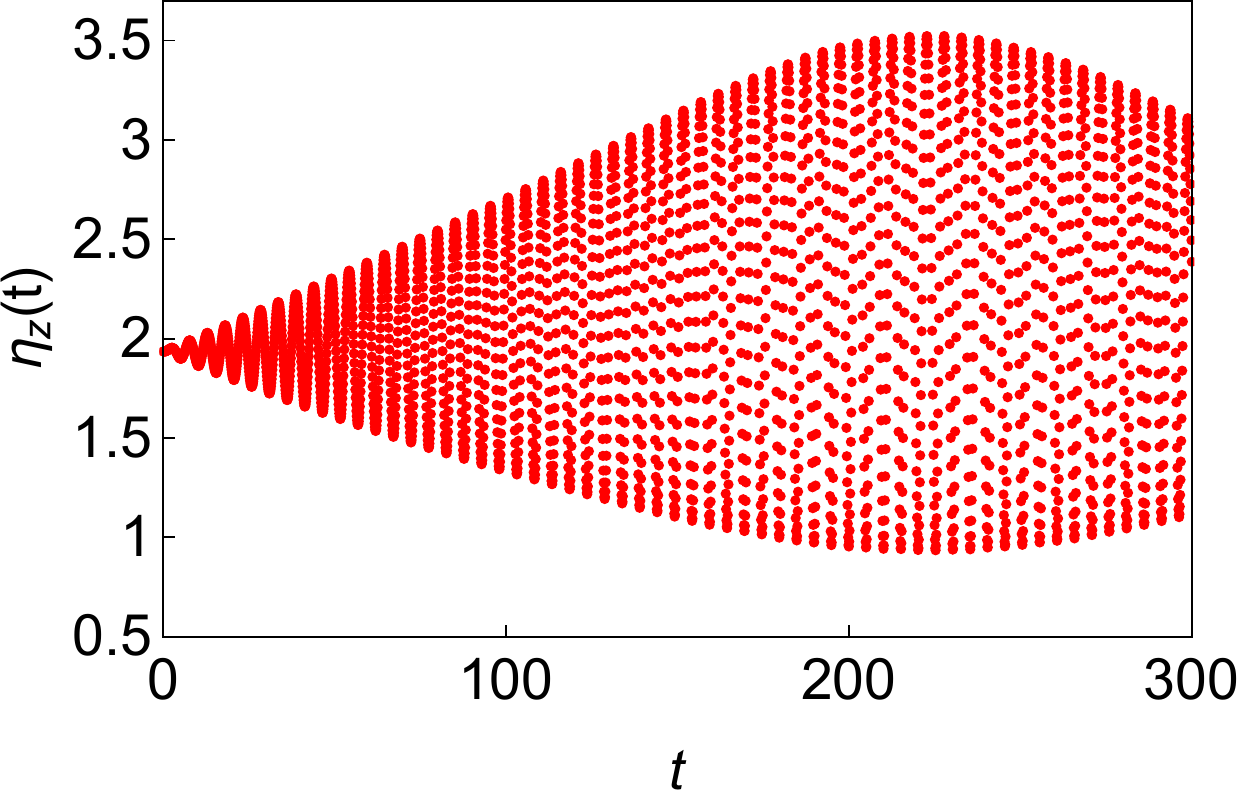}
	\caption{(Colour online) Dynamics $\eta_{\rho}(t)$ and $\eta_z(t)$ in unit of $\omega_{\rho}t$ vs $t$ for $\lambda=0.7$, $\delta=0.1$, $\mathcal{P}=10$, $q=0.01$, $\Omega=1.2\approx \omega_Q$.}
	\label{fig:dynamics}
\end{figure}

Large amplitude oscillations are present for $\Omega=1.2$ close to $\omega_Q$ as shown in figure~\ref{fig:dynamics} which is an
example close to a resonance. Therefore, an analytical
description would be necessary to prove that it is not only some numerical artefact of our procedure, but real nonlinear
effect is induced via anharmonic distortion in the physical system.
We consider equations~(\ref{eq:Rr}) and~(\ref{eq:Rz}) for small driving $q$, with the following
initial conditions:
 \begin{eqnarray}
&&\eta_{\rho}(0)=\eta_{\rho_0}\,, \qquad \eta_{z}(0)=\eta_{z_0}, \nonumber\\
&&\dot \eta_{\rho}(0)=\dot \eta_{z}(0)=0\,,
\label{eq:intialCon}
 \end{eqnarray}
then, the eigenmode frequencies equation~(\ref{eq:frequencyBQ}) are expected not to be excited, thus
simplifying the behavior of the system and allowing the application of the standard Poincar\'e--Lindstedt method to perform perturbative expansions for small $q$~\cite{Hamid1,Hamid2}:
\begin{eqnarray}
&&\eta_{\rho}(t)=\eta_{\rho_0}+q \ \eta_{\rho1}(t)+q^2 \ \eta_{\rho2}(t)+q^3\ \eta_{\rho3}(t)+... \ ,\nonumber\\
&&\eta_{z}(t)=\eta_{z_0}+q \ \eta_{z1}(t)+q^2\ \eta_{z2}(t)+q^3\ \eta_{z3}(t)+...\ .
\label{eq:perturbative+approach}
\end{eqnarray}
By inserting equation~(\ref{eq:perturbative+approach}) into equations~(\ref{eq:Rr}) and~(\ref{eq:Rz}), we find a system of decoupled linear
differential equations
\begin{equation}
\left(\begin{array}{c} \ddot \eta_{\rho n}(t)\\ \ddot \eta_{zn}(t) \end{array}\right) + \left(\begin{array}{cc} \alpha_1 & \alpha_2\\ 2\alpha_2 & \alpha_3 \end{array}\right)  \left(\begin{array}{c} \eta_{\rho n}(t)\\ \eta_{zn}(t) \end{array}\right)
+ \left(\begin{array}{c}  k_{\rho n}(t)\\ k_{zn}(t) \end{array}\right)=0.
\end{equation}
The terms $k_{\rho n}$ and $k_{zn}$ rely on the lower order solutions $\eta_{\rho i}$ and $\eta_{zi}$, such as $i<n$. At $n=1$, $k_{\rho 1}$ and $k_{z 1}$ have the form as $8 \eta_{\rho_0}^3 \sin(\Omega \ t)$ and $0$, respectively.
${\bf \eta}_1=\left(\begin{array}{c} \eta_{\rho 1}\\ \eta_{z1} \end{array}\right) $ is found and this solution is used to calculate ${\bf \eta}_2$. At the $n$-th level of this iterative procedure, we use the initial conditions
equations~(\ref{eq:intialCon}).
This is a common system of linear differential equations describing a well-known forced harmonic oscillator. Its solution possesses precisely two modes, $\omega=\omega_{Q,B}$ and $\omega=\Omega$, and these modes are denoted as basic modes. The main resonance is located at $\Omega=\omega_{Q,B}$. In the case of $\Omega \neq \omega_{Q,B}$, the basic modes in figure~\ref{fig:spectrum} are the most prominent ones in the spectrum, indicating that approximation $\eta_1(t)$ contains important physics of equations~(\ref{eq:Rr}) and~(\ref{eq:Rz}).

\begin{figure}[htb]
	\centering
	\includegraphics[width=6cm]{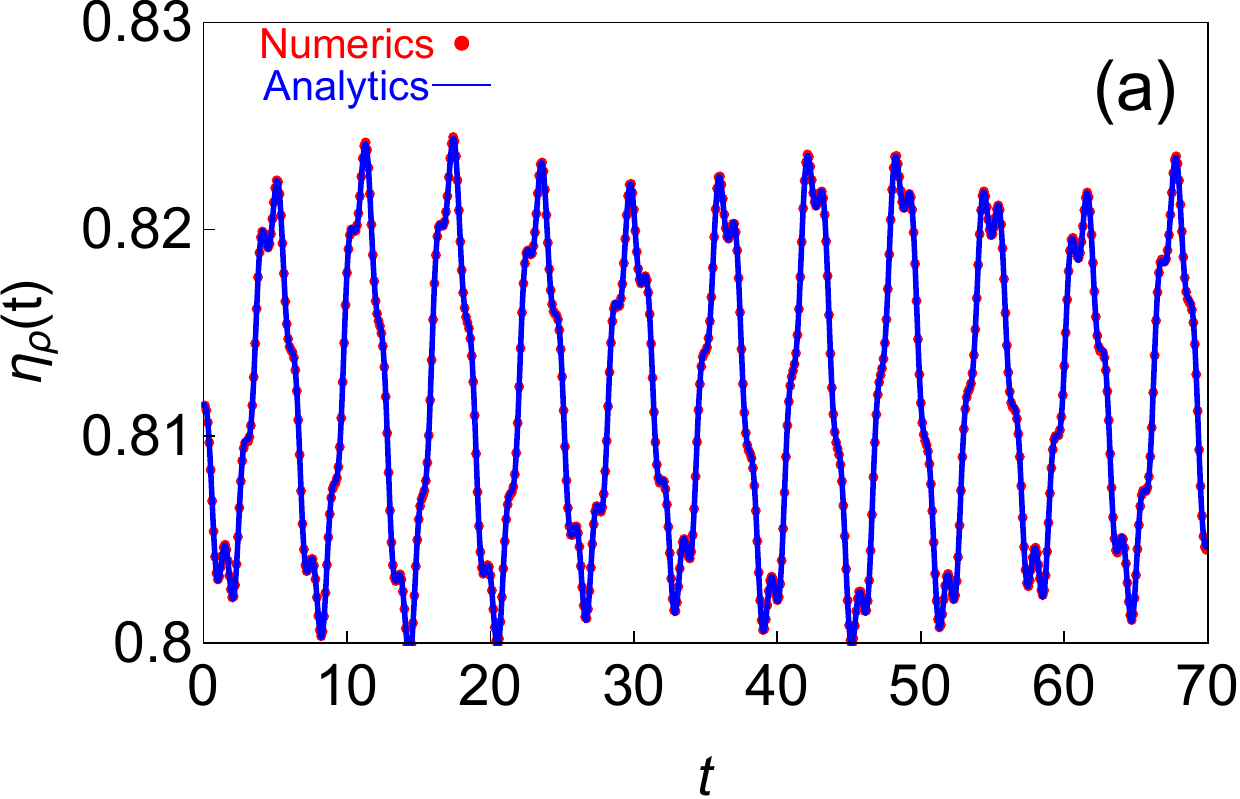}
	\includegraphics[width=6cm]{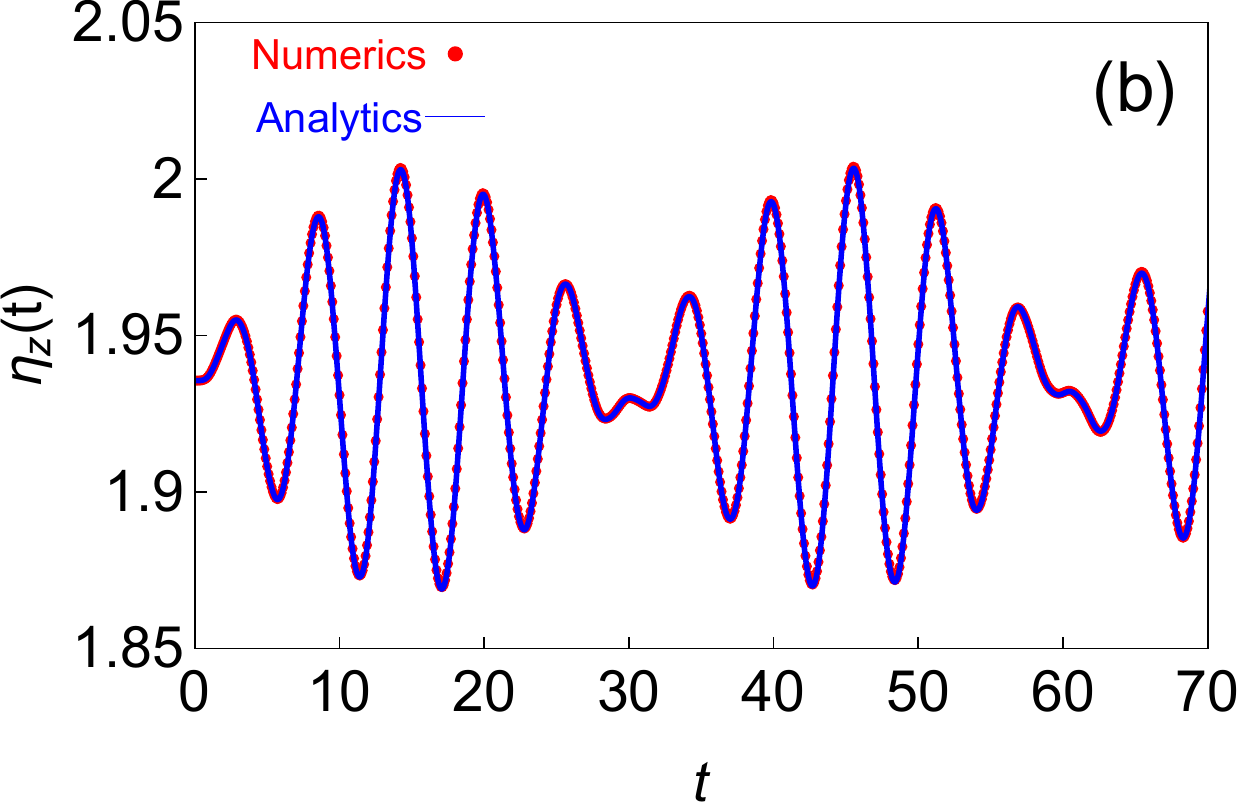}
	\caption{(Colour online) Dynamics $\eta_{\rho}(t)$ and $\eta_z(t)$ in unit of $\omega_{\rho}t$ vs $t$ for $\lambda=0.7$, $\delta=0.1$, $\mathcal{P}=10$, $q=0.01$, $\Omega=0.4$.}
	\label{fig:dynamics1}
\end{figure}

Figure \ref{fig:dynamics1}, presents an example of the condensate dynamics as shown for $\lambda=0.7$, $\delta=0.1$, $\mathcal{P}=10$, $q=0.01$, $\Omega=0.4$ by plotting numerical solutions of equations~(\ref{eq:Rr}) and~(\ref{eq:Rz}) with the initial conditions equation~(\ref{eq:intialCon})   together with the analytical result which is based on perturbation theory in $q$. A good agreement of analytic and numerical results was obtained for long extension  times due to figure~\ref{fig:dynamics1}, which leads to a good accuracy of the Gaussian approximation for determining the oscillation mode frequencies of the excited modes.

\section{Stability analysis}
\label{sec:stability}

\begin{figure}[htp]
\centering
\includegraphics[width=6cm]{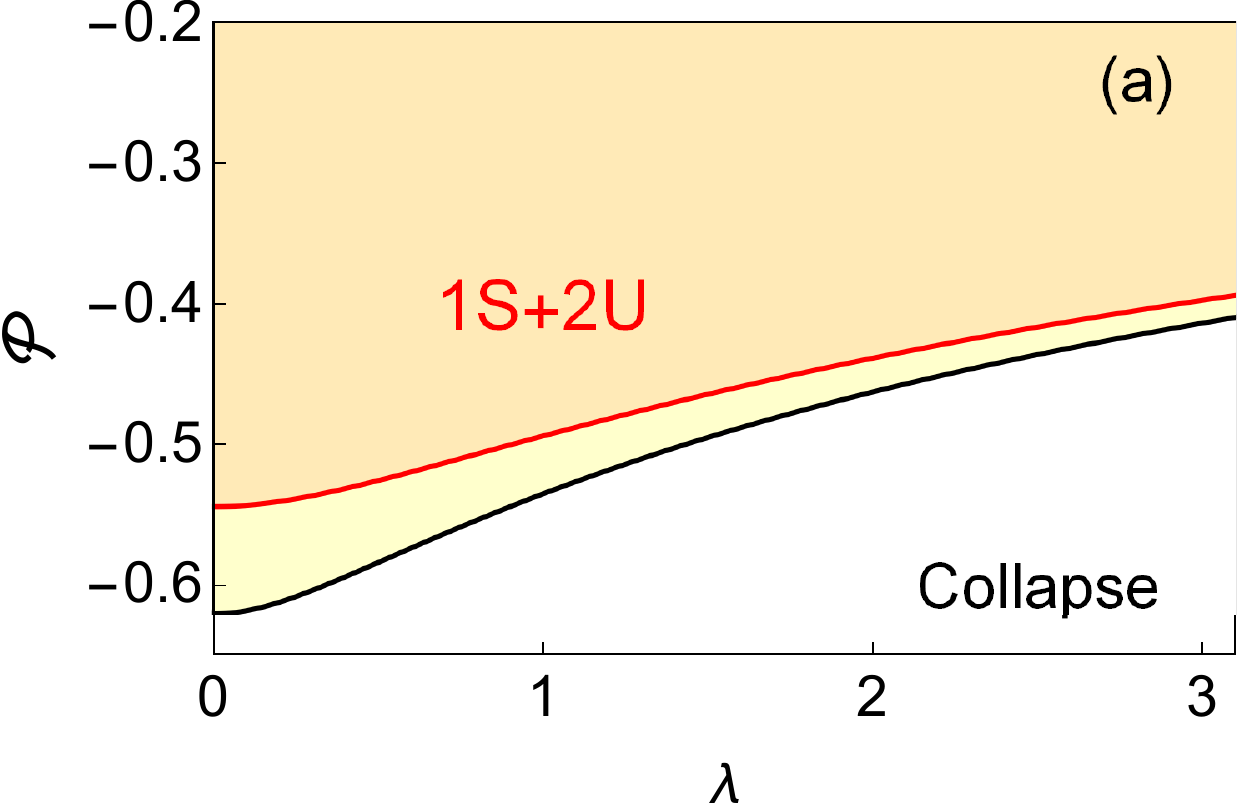}
\includegraphics[width=6cm]{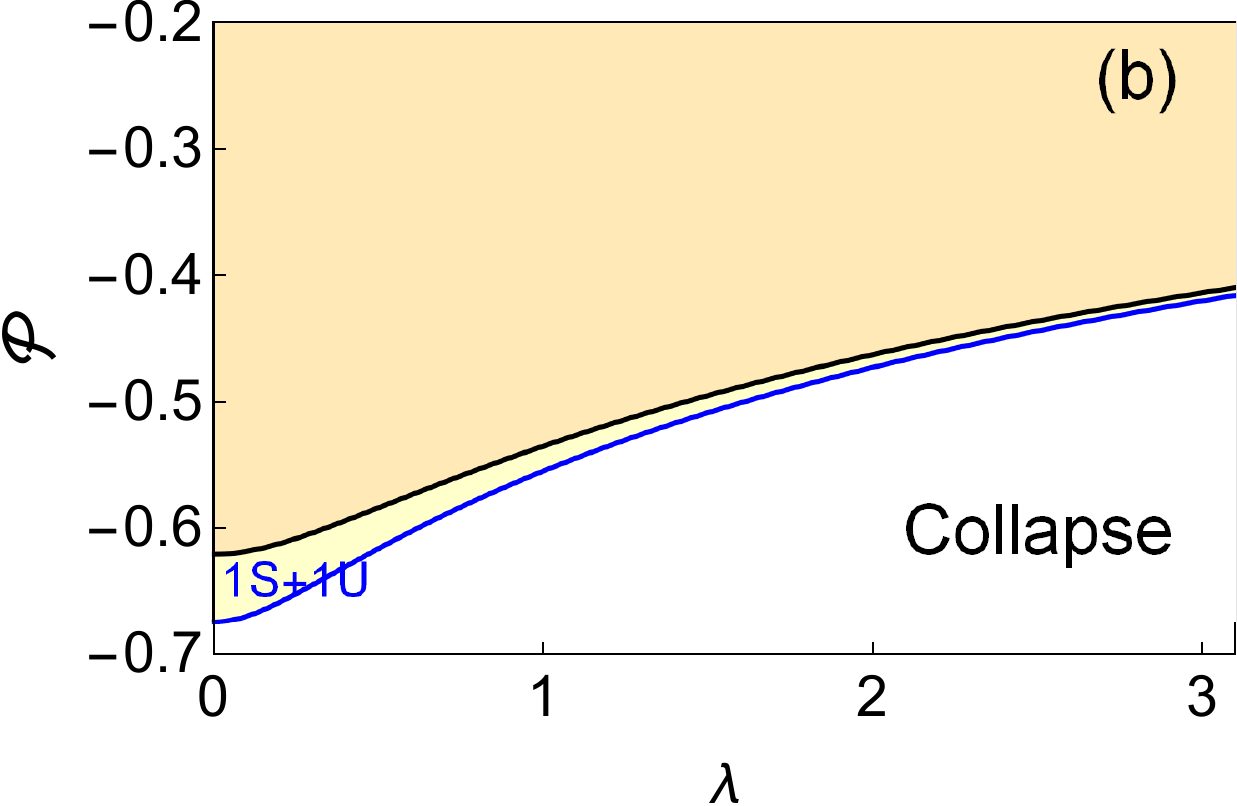}
\caption{(Colour online) $\lambda$-$\mathcal{P}$ stability scheme for $\delta=0$ (black curve). Below the black curve the condensate is collapse. The condensate has two solutions for $\mathcal{P}<0$ (above the black curve). One is stable, whereas the other is unstable. The condensate has one stable solution  (1S) for $\mathcal{P}\geqslant 0$. (a)~$\lambda$-$\mathcal{P}$ stability scheme with anharmonic distortion $\delta=0.002$ (red curve). (b) $\lambda$-$\mathcal{P}$ stability scheme for $\delta=-0.005$ (blue curve).}
\label{fig:stabilites}
\end{figure}

The stability of the BEC is studied in this section for systems with small interaction strengths at zero temperature in an external axially symmetric plus anharmonic distortion trap. BECs in systems with attractive interaction (i.e., $\mathcal{P}<0$) are unstable for a large enough negative value of $\mathcal{P}$~\cite{pitaevskii+stringari,CJPethick}.
As we demonstrate, stable condensate is above the stability black curve in figure~\ref{fig:stabilites} for two-body interaction, while the existence of a small repulsive (attractive) anharmonic distortion leads to a destabilisation (stabilisation) of the condensate. To discuss in detail the influences of anharmonic distortion of BEC stability systems, several interesting cases are examined: repulsive and attractive anharmonic distortion within two-body interactions. If the equations~(\ref{eq:Rr}) and~(\ref{eq:Rz}) have real
and limited solutions in the proximity of positive equilibrium condensate widths, then the condensate is assumed stable, otherwise the system is unstable.

This is identical to implementing a linear stability
analysis and calculating the stability of positive equilibrium widths by
testing the eigenmode frequencies equation~(\ref{eq:frequencyBQ}). The solution is
stable if the eigenmode frequencies are real.
Black curve in figure~\ref{fig:stabilites} refers to the critical stability line versus $\lambda$ for inter-atomic interaction strength (i.e., $\delta=0$). There are no stable solutions under the black line (unstable condensate). For $\mathcal{P}<0$ (over black curve), the condensate has two solutions (i.e., one unstable and one stable), while one stable
solution for $\mathcal{P}\geqslant 0$. \

The case of a small repulsive anharmonic distortion $\delta=0.002$ and $\mathcal{P}<0$ is assumed, and the outcomes of stability analysis are completely different. The equation has three solutions, one is stable, while the other two solutions are unstable, as shown in figure~\ref{fig:stabilites}~(a) (red curve). Below the red curve, the condensate is unstable. On the other hand, if we assume an attractive anharmonic distortion $\delta=-0.005$ figure~\ref{fig:stabilites}~(b) (blue curve), the stability analysis of the condensate has two solutions, one is stable and the other one is unstable.

\section{Conclusions}
\label{sec:con}
The dynamics of a BEC and their stability analysis is studied in an axially symmetric plus anharmonic distortion trap. According to variational approaches, the time-dependent GP equation is demonstrated and the equation for the dynamics of the condensate widths is obtained. The resonance, mode coupling, and nonlinear analysis by modulating the anharmonicity of the trapping potential is discussed. It is shown how a small repulsive (attractive) anharmonic distortion affects the system of attractive interaction strength, where the stability region is reduced (extended) and the condensate decreases (increases) the critical particle number. Furthermore, our future step is to study
the effects of the modulation for an anharmonic distortion within negative s-wave scattering length which amounts to
the effect of parametric stabilization when some partially unstable systems can be stabilized
by periodic driving.

\ukrainianpart

\title{Динаміка Бозе--Ейнштейнівських конденсатів у ангармонічних пастках}

\author{Х. Аль-Джибурі}
\address{Науковий колледж університету Аль-Кадісія, Аль-Діванія, 58002, Аль-Кадісія, Ірак}
\makeukrtitle

\begin{abstract}
	Досліджено динаміку слабовзаємодіючих тривимірних Бозе--Ейнштейнівських конденсатів (БЕК), утримуваних за допомогою зовнішнього аксіально-симетричного потенціалу при наявності ангармонічної складової. У рамках варіаційного підходу з використанням залежного від часу рівняння Гросса--Пітаєвського отримана система рівнянь для визначення ширини області конденсату. Модулюючи ангармонічне спотворення потенціалу пастки, було досліджено чисельно та проілюстровано аналітично такі нелінійні особливості, як взаємодія між коливними модами та резонансами. Крім того, досліджено стійкість притягальної взаємодії БЕК як у випадку відштовхувальної, так і притягальної складових ангармонічного потенціалу.	Показано, що наявність незначної відштовхувальної чи притягальної частини ангармонічного потенціалу ефективно звужує (розширює) область стійкості конденсату, оскільки зменшує (збільшує) критичну кількість атомів у потенціальній пастці.
	\keywords Бозе--Ейнштейнівські конденсати, рівняння Гросса--Пітаєвського, ангармонічна пастка
\end{abstract}

\lastpage
\end{document}